%% file: project.tex
\documentclass[conference,flushend]{iaria}
\pdfoutput=1 

\usepackage[babel=true,english=american]{csquotes}
\usepackage[english]{babel} 

\usepackage[shortcuts]{extdash} 

\usepackage[alwaysadjust]{paralist}
\usepackage[ruled]{algorithm2e}
\makeatletter
\let\c@algocf\c@figure

\SetAlgorithmName{Figure}{figure}{List of Figures}
\makeatother
\usepackage{subcaption}
\usepackage{graphicx}

\usepackage{listings}

\usepackage[
babel=true, 
expansion=alltext,
protrusion=alltext-nott, 
nopatch=eqnum, 
final 
]{microtype}

\usepackage{fontawesome} 
\usepackage{relsize}     
\usepackage{lipsum}      
\usepackage{xurl}       

\newcommand{\papertitle}{QuaQue: Design and SQL Implementation of Condensed Algebra for Concurrent Versioning of Knowledge Graphs}
\title{\papertitle}

\author{
    \IEEEauthorblockN{Jey Puget Gil, \orcidlink{0009-0006-6198-7488} Emmanuel Coquery, \orcidlink{0000-0001-7015-5604} John Samuel, \orcidlink{0000-0001-8721-7007} Gilles Gesquière \orcidlink{0000-0001-7088-1067}}
    \IEEEauthorblockA{%
        Universite Claude Bernard Lyon 1, CNRS, INSA Lyon, Université Lumière Lyon 2,\\
        Ecole Centrale de Lyon, CPE Lyon, LIRIS, UMR 5205\\
        Villeurbanne, France\\
        e-mail: {\tt$\lbrace$jey.puget-gil\,|\,emmanuel.coquery\,|\,john.samuel\,|\,gilles.gesquiere$\rbrace$@liris.cnrs.fr}
    }
}

\begin{document}

\include{content/style-listings}

\maketitle

\input{content/abstract}

\begin{IEEEkeywords}
    SQL; algebra; translation; concurrent versioning; relational.
\end{IEEEkeywords}

\input{content/chapters/introduction}
\input{content/chapters/state-of-the-art}

\input{content/chapters/quaque}
\input{content/chapters/benchmark}

\input{content/chapters/conclusion}

\input{content/acknowledments}

\printbibliography

\input{content/appendix}

\end{document}

%% file: content/style-listings.tex
\lstdefinelanguage{SPARQL}{
  basicstyle=\small\ttfamily,
  columns=fullflexible,
  breaklines=true,
  sensitive=true,
  tabsize = 2,
  showstringspaces=false,
  morecomment=[l][\color{green}]{\#},       
  morecomment=[n][\color{red}]{<}{>}, 
  morestring=[b][\color{orange}]{\"},  
  keywordsprefix=?,
  classoffset=0,
  keywordstyle=\color{darkgray},
  morekeywords={},
  classoffset=1,
  keywordstyle=\color{purple},
  morekeywords={rdf,vers},
  classoffset=2,
  keywordstyle=\color{blue},
  morekeywords={
    SELECT,CONSTRUCT,DESCRIBE,ASK,WHERE,FROM,NAMED,PREFIX,BASE,OPTIONAL,
    FILTER,GRAPH,LIMIT,OFFSET,SERVICE,UNION,EXISTS,NOT,BINDINGS,MINUS,a
  }
}

\lstdefinelanguage{SQL}{
  basicstyle=\small\ttfamily,
  columns=fullflexible,
  breaklines=true,
  sensitive=true,
  tabsize = 2,
  showstringspaces=false,
  morecomment=[l][\color{green}]{--},       
  morecomment=[n][\color{red}]{/*}{*/}, 
  morestring=[b][\color{orange}]{\'},  
  classoffset=0,
  keywordstyle=\color{darkgray},
  morekeywords={},
  classoffset=1,
  keywordstyle=\color{purple},
  morekeywords={},
  classoffset=2,
  keywordstyle=\color{blue},
  morekeywords={
    SELECT,FROM,WHERE,JOIN,INNER,LEFT,RIGHT,FULL,ON,AS,DISTINCT,ALL,AND,OR,NOT,
    IN,IS,NULL,LIKE,GROUP,BY,HAVING,ORDER,DESC,ASC,LIMIT,OFFSET,UNION,EXCEPT,
    INTERSECT,INSERT,INTO,VALUES,UPDATE,SET,DELETE,CREATE,TABLE,PRIMARY,KEY,
    FOREIGN,REFERENCES,DROP,ALTER,ADD,COLUMN
  }
}

%% file: content/abstract.tex
\begin{abstract}
    The management of versioned knowledge graphs presents significant challenges, particularly in querying data across multiple versions efficiently.
    This paper introduces QuaQue, a key component of the ConVer-G system, which addresses this challenge by translating SPARQL (SPARQL Protocol and RDF Query Language) queries into SQL (Structured Query Language).
    QuaQue leverages a novel condensed algebra to operate on a relational model where versioning information is compactly stored using bitstrings.
    This approach allows for efficient querying of concurrent versions of knowledge graphs within a standard relational database system.
    We present the key concepts of our condensed algebra, detail the translation process from SPARQL algebra to SQL, and provide a comparative benchmark against a native RDF (Resource Description Framework) triple store, demonstrating the viability and performance benefits of our approach.
\end{abstract}

%% file: content/chapters/introduction.tex
\section{Introduction}
The representation of complex, evolving information has driven the widespread adoption of Knowledge Graphs (KGs) in both industry and academia~\cite{hogan2021knowledge}.
However, as KGs move from static repositories to dynamic assets, there is a critical need for robust \textit{concurrent versioning systems}.
In domains, such as urban planning—where datasets undergo parallel modifications by multiple stakeholders—a linear history is insufficient.
Systems must support branching, merging, and the analysis of concurrent states without excessive data redundancy.

While the Resource Description Framework and SPARQL are the de facto standards for KGs, they were primarily designed for static or monotonically increasing datasets.
Native support for versioning remains a challenge.
Standard approaches often rely on Named Graphs~\cite{carroll2005named} to isolate versions.
While accessing a single version remains efficient, this strategy leads to significant data duplication and increased query latency when querying across multiple versions.
Consequently, efficiently querying across multiple versions remains an open problem in database research.

The ConVer-G project~\cite{gil2024conver} addresses this by bridging the gap between graph versioning requirements and the mature optimization capabilities of Relational Database Management Systems (RDBMSs).
At the core of ConVer-G is \textbf{QuaQue}, a system that translates SPARQL queries into SQL that exploits efficient bitwise operations for version filtering.

QuaQue leverages a \textit{condensed relational model}.
In RDF, a \textit{quad} is a tuple that extends the standard triple with a graph identifier.
We associate every quad with a \textit{bitstring}, where each bit represents the validity of the quad in a specific version.
This allows us to push version-filtering logic down to the RDBMS engine using efficient bitwise operations, significantly reducing the I/O overhead typically associated with multi-version queries.

This paper details the design and implementation of QuaQue.
We introduce a \textit{condensed algebra}—an extension of relational algebra tailored for bitstring-annotated relations—and define its translation to SQL.
This enables the execution of complex graph pattern matching on standard PostgreSQL instances.

Our specific contributions are:
\begin{itemize}
    \item A novel \textbf{condensed relational model} utilizing bitstrings for the storage of concurrent KG versions.
    \item The implementation of a \textbf{Condensed Algebra} and the QuaQue translator which maps SPARQL algebra to SQL.
    \item A comparative benchmark demonstrating that QuaQue outperforms a native RDF triple store (Apache Jena) in multi-version query scenarios.
    \item A fully reproducible approach, with the ConVer-G tool and the benchmark framework publicly available as open-source software.
\end{itemize}

The remainder of this paper is organized as follows.
Section 2 presents the state of the art.
Section 3 details the design of the QuaQue system and the condensed relational model.
Section 4 presents the experimental evaluation and benchmark results.
Finally, Section 5 concludes the paper and outlines future work.

%% file: content/chapters/state-of-the-art.tex
\section{State of the Art}
The challenge of querying versioned data has been a long-standing topic in database research~\cite{polleres2023does}.
The relational model~\cite{codd1970relational}, introduced by Codd, provides a foundation with relational algebra and calculus as powerful query languages.
The expressive power of these languages has been a central theme of research, with extensions proposed to handle more complex queries, such as those involving recursion~\cite{roth1988extended, agrawal2002alpha} and aggregate functions~\cite{ozsoyouglu1987extending}.
A key milestone in bridging the gap between high-level query languages and efficient execution is the work of Ullman~\cite{ullman1985implementation}, which systematically addresses the translation of relational algebra and calculus queries into implementations, laying the groundwork for query processing and optimization.
This connection is important, as the development of new algebras or extensions—such as those needed for versioned or condensed data—must ultimately be supported by practical translation and execution strategies.
Recent surveys, such as Hofer et al.~\cite{hofer2023construction}, Jiang et al.~\cite{jiang2023evolution, abbas2020knowledge}, and Ji et al.~\cite{ji2021survey} provide a comprehensive overview of the current state and challenges in knowledge graph construction, highlighting the increasing complexity of managing evolving and versioned knowledge graphs.
They emphasize the need for scalable and efficient methods for both constructing and maintaining knowledge graphs, especially as these graphs become larger and more dynamic.
This underscores the importance of advanced versioning and querying mechanisms to support the evolving requirements of knowledge graph applications.
A notable example of leveraging versioned knowledge graphs in practice is the work by Gonzalez-Hevia and Gayo-Avello~\cite{gonzalez2022leveraging}, who utilize Wikidata's edit history for knowledge graph refinement tasks.
Their approach demonstrates the value of exploiting historical edit information to improve the quality and reliability of knowledge graphs, further motivating the need for efficient storage and querying of versioned data.

Recent work by Zhong et al.~\cite{zhong2024data} further illustrates the importance of knowledge graphs in real-world applications, specifically in the domain of intelligent audit.
Their study discusses both the opportunities and challenges of applying knowledge graphs to extract insights from complex, evolving data sources.
This highlights the direct link between the need for advanced versioning and querying mechanisms—such as those discussed in this section—and the practical requirements of domains where data evolution and efficient cross-version analysis are critical for generating reliable insights.

\subsection{Extending Relational Algebra}
Relational algebra has been extended with techniques, such as rewriting queries with arbitrary aggregation functions using views, as discussed by Cohen et al.~\cite{cohen2006rewriting}.
The extensibility of relational algebra has also been a subject of research, particularly in the context of supporting new data types and operations.
Haas et al.~\cite{haas1989extensible} proposed an extensible query processor architecture that allows the integration of user-defined types and functions into the relational algebra framework.
This extensibility adapts relational systems to various application domains, enabling the seamless incorporation of specialized operators and data structures without sacrificing the benefits of a declarative query language.

\subsection{Relational Algebra in Query Languages}
The principles of relational algebra have been foundational to the design of numerous query languages.
SQL, the de facto standard for relational databases, is based on a tuple relational calculus, which is equivalent in expressive power to relational algebra~\cite{codd1972relational}.
The translation of SPARQL~\cite{cyganiak2005relational}, the standard query language for RDF, to SQL has been a topic of interest for leveraging the performance and scalability of relational databases for semantic web data.
Several systems, such as Ontop~\cite{calvanese2016ontop}, have explored this translation, often relying on mappings between RDF and relational schemas.

Relational algebras are also applied outside the context of traditional databases.
In qualitative spatial and temporal reasoning~\cite{duntsch2005relation, li2021temporal}, relation algebras serve as formal tools for modeling and inferring relationships between spatial or temporal entities.
For instance, Allen's interval algebra~\cite{allen1983maintaining} provides a calculus for reasoning about temporal intervals, while the Region Connection Calculus (RCC)~\cite{cohn1997qualitative} is used for spatial reasoning.
These algebras provide a formal, equational framework for deriving new knowledge from a set of base relations.

\subsection{Semantic Versioned Querying: The Fundamentals}
A contribution to the field of versioned querying for knowledge graphs is presented by Taelman et al.~\cite{taelman2018fundamentals}.
Their work systematically investigates the requirements and challenges of querying versioned semantic data, focusing on the formalization of versioned query semantics and the practical implications for query processing.

Taelman et al.\ introduce a formal framework for semantic versioned querying, distinguishing between different types of versioned queries, such as snapshot queries (retrieving data as it existed at a specific version), longitudinal queries (tracking the evolution of data across versions), and difference queries (identifying changes between versions).
They emphasize the importance of clearly defined semantics for each query type, as ambiguity can lead to inconsistent or unintuitive results.

A key insight from their work is the need for query languages and systems to natively support version-aware operations, rather than treating versioning as an afterthought or external feature.
They propose extensions to SPARQL that allow users to specify version constraints directly within queries, enabling more expressive and precise retrieval of historical or evolving data.

Overall, the work of Taelman et al.\ provides a foundational perspective on the semantics of querying versioned knowledge graphs.
Their formalization of query types has been influential in our work, serving as a guideline for the capabilities our system aims to support.

\subsection{Versioning Models for Evolving Data}
Foundational models for data evolution stem from Software Configuration Management (SCM)~\cite{conradi1998version}, which distinguishes between sequential revisions and parallel variants.
This distinction applies to knowledge graph versioning, where evolution involves both temporal changes and concurrent viewpoints~\cite{gil2025condensed}.
SCM also differentiates between state-based (snapshots) and change-based (deltas) models~\cite{conradi1998version}.
In Model-Driven Engineering (MDE), these concepts extend to graph structures, where revisions are defined as sequences of atomic graph modifications~\cite{taentzer2014fundamental}.
Formalizing changes as graph operations enables structured reasoning about conflicts and merging \cite{graube2014r43ples}, providing a theoretical basis for versioned query algebras.

Several strategies for storing versioned RDF data have emerged, balancing storage efficiency and query performance:

\subsubsection{Independent Copies (IC)}
The IC or snapshot approach stores each version as a full copy~\cite{volkel2005semversion}.
While efficient for single-version querying (Version Materialization), it suffers from high storage redundancy.

\subsubsection{Change-Based (CB)}
CB or delta-based versioning stores a base version and subsequent changes.
This is space-efficient but requires costly reconstruction for querying.
Tools, such as R43ples~\cite{graube2014r43ples}, R\&WBase~\cite{vander2013r}, and Stardog, follow this paradigm.

\subsubsection{Timestamp-Based (TB)}
TB approaches annotate triples with validity intervals, facilitating ``time-travel'' queries.
While suited for linear evolution, supporting branching adds complexity.
Tools adopting this strategy include ConVer-G~\cite{gil2024convergconcurrentversioningknowledge}, Drydra~\cite{anderson2016transaction}, RDF-TX~\cite{gao2016rdf}, v-RDFCSA~\cite{7786197}, and x-RDF-3X~\cite{10.14778/1920841.1920877}, often drawing on temporal database concepts~\cite{10.1145/2380776.2380786}.

\subsubsection{Fragment-Based (FB)}
FB or hybrid approaches partition the graph into independently versioned fragments, balancing storage and query performance.
However, managing dependencies between fragments is complex.
QuitStore~\cite{Arndt201929} implements this by versioning modified files in a Git repository.

\subsubsection{Graph Compression}
Graph compression techniques, such as HDT (Header, Dictionary, Triples)~\cite{fernandez2013binary}, offer another perspective on efficient RDF storage.
HDT is a binary format that achieves high compression ratios by utilizing dictionary encoding for RDF terms and a compact bitmap-based structure for the graph topology.
Crucially, HDT files are designed to be queryable directly without decompression, providing excellent read performance for static datasets.
However, the primary limitation of HDT and similar compression-centric approaches in the context of versioning is their static nature.
They are optimized for read-only scenarios; modifying the data typically requires a computationally expensive reconstruction of the entire file.
While one could theoretically store each version as a separate HDT file (effectively an optimized IC approach), this does not inherently solve the redundancy problem for small incremental changes, nor does it facilitate efficient cross-version querying or branching/merging operations.
Therefore, while compression is a valuable component of storage optimization, it does not by itself constitute a complete versioning strategy for dynamic, evolving knowledge graphs.

\subsubsection{Limitation}
Git-based solutions, such as R\&WBase~\cite{vander2013r} and QuitStore, require explicit checkout to query a version, hindering concurrent cross-version analysis~\cite{gil2024conver}.
Existing strategies struggle to balance storage efficiency and query performance for such analysis.

\subsection{Summary: The Case for a Condensed TB Representation and Algebra}
Current versioning strategies face a trade-off between storage and query efficiency, often lacking support for concurrent cross-version analysis~\cite{cuevas2020versioned}.
A \textbf{condensed TB representation} addresses this by storing unique quads annotated with version validity, as seen in ConVer-G~\cite{gil2024conver}.
To exploit this model, a condensed algebra is needed to define operators on version-annotated structures, similar to specialized algebras, such as Knowledgebra~\cite{yang2022knowledgebra}.
Finally, an algebra-to-SQL translator (QuaQue) bridges the gap to efficient execution on relational systems~\cite{gil2025condensed}.

%% file: content/chapters/quaque.tex
\section{Design}
The QuaQue component of the ConVer-G system is a SPARQL-to-SQL translator designed to query a condensed relational model of versioned RDF data.
Our approach is motivated by the need for efficient cross-version queries, which are cumbersome and inefficient with traditional triple-store-based versioning methods that replicate data for each version.

\subsection{Condensed Relational Model}
Our condensed model represents versioned RDF data in a relational database management system.
We chose PostgreSQL for our implementation because it provides native support for bit string data types and efficient bitwise operations.
This representation avoids storing duplicate quads that exist across multiple versions.
The interaction between the SPARQL translation and this model is depicted in Figure~\ref{fig:translation-overview}.
Versioned quad table, which stores quads (subject, predicate, object, named graph) along with a validity bitstring.
Each bit in the validity string corresponds to a specific version, and a '1' at a given position indicates that the quad is present in that version.
The structure of our condensed relational model is illustrated in Figure~\ref{fig:conceptual-model} and described as follows:

\begin{figure}[h]
    \centering
    \includegraphics[width=\textwidth]{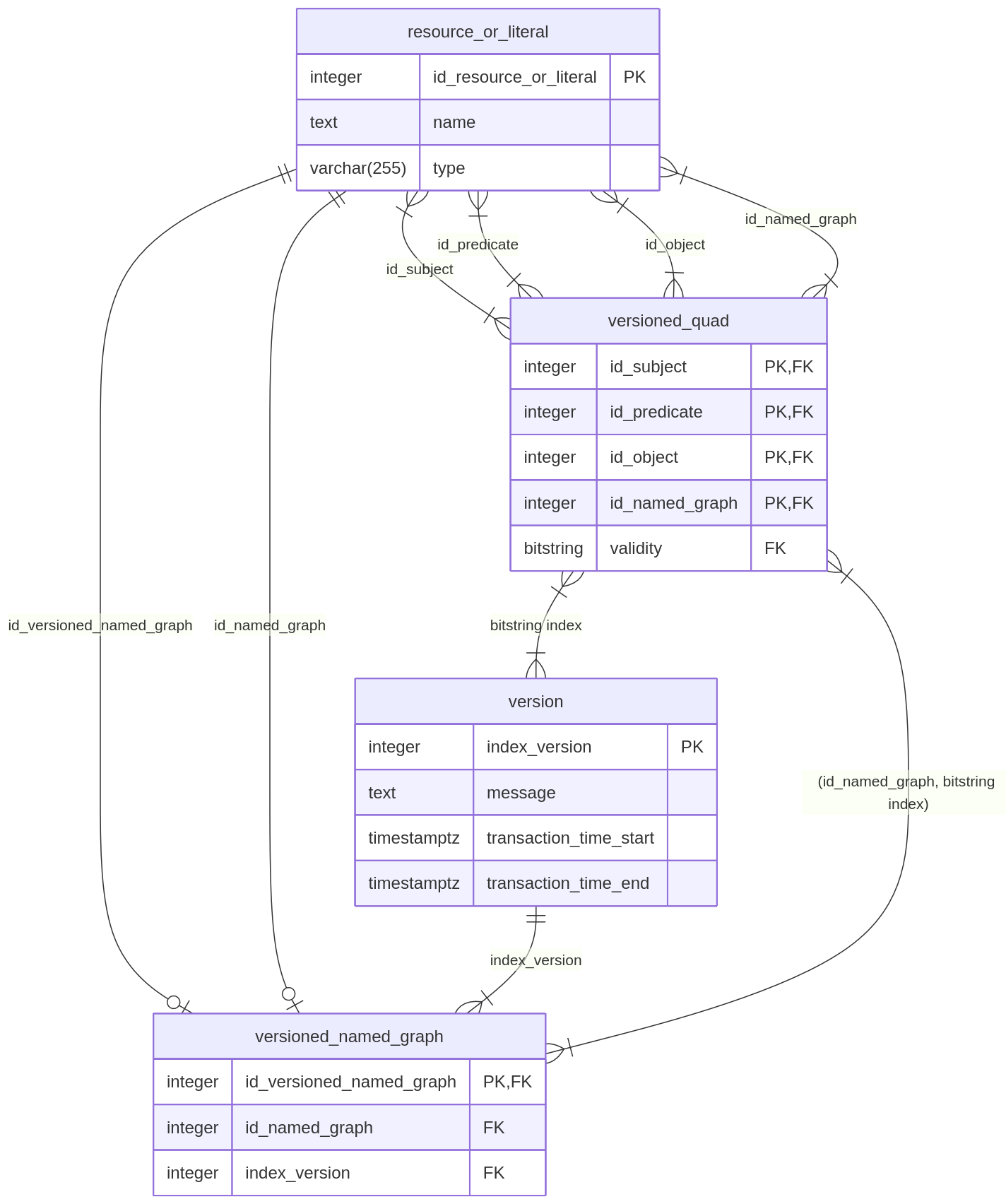}
    \caption{Relational model of the condensed representation for versioned RDF data.}
    \label{fig:conceptual-model}
\end{figure}

The schema comprises the following relations:
\begin{itemize}
    \item \texttt{versioned\_quad}: This is the central table, storing unique quads (subject, predicate, object, named graph) as integer identifiers. Each entry includes a validity bitstring that indicates the versions in which the quad is present. This design minimizes redundancy by storing each unique quad only once, regardless of how many versions it appears in. An example is provided in Table~\ref{tab:condensed-dataset}.
    \item \texttt{resource\_or\_literal}: This table acts as a dictionary, mapping RDF terms (URIs and literals) to the integer identifiers used in other tables. This practice, known as dictionary encoding, optimizes storage and join performance by replacing long strings with compact integers. See Table~\ref{tab:dictionary-dataset} in the Appendix for an example.
    \item \texttt{version}: This table holds metadata specific to each version, such as its creation timestamp or a descriptive label. Separating version metadata allows for efficient retrieval of version-specific information without scanning the quad data.
    \item \texttt{versioned\_named\_graph}: This table links named graphs to the versions they belong to. This separates the association between graphs and versions from the quad data, adhering to database normalization principles to support scenarios where named graphs evolve independently across versions.
    \item \texttt{metadata}: This table stores additional metadata, which can be user-defined, about versions and named graphs. This flexibility accommodates diverse application requirements for tracking contextual information.
\end{itemize}

This schema design ensures the storage and retrieval of versioned RDF data while maintaining the flexibility required for diverse application scenarios.

\section{Development}

\subsection{Condensed Algebra and SQL Translation}
\begin{figure}
    \centering
    \includegraphics[width=0.95\textwidth]{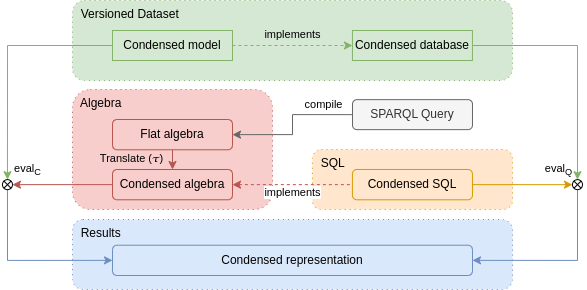}
    \caption{Overview of the SPARQL to SQL translation process in QuaQue.}
    \label{fig:translation-overview}
\end{figure}

The QuaQue component translates SPARQL queries into SQL by first converting the SPARQL query into a SPARQL algebra expression, and then mapping the SPARQL algebra operators to SQL operations on our condensed model, as shown in Figure~\ref{fig:translation-overview}.

The core of our approach lies in how we handle the validity bitstring.
For operations that combine quads, such as joins, we use bitwise operations on the validity bitstrings.
For example, a join between two quad patterns corresponds to a bitwise AND operation on their validity bitstrings.
This allows us to efficiently determine the versions in which the joined pattern is valid.

The translation process can be summarized as follows (see Figure~\ref{fig:translation-overview}):
\begin{itemize}
    \item \textbf{SPARQL to SPARQL Algebra:} The incoming SPARQL query is parsed into a SPARQL algebra expression tree using Apache Jena.
    \item \textbf{SPARQL Algebra to Condensed SQL:} The SPARQLtoSQLTranslator traverses the algebra tree and, for each operator, generates a corresponding SQL query fragment.
          The QuadPatternSQLOperator handles the base case of translating a quad pattern into a SQL query on the \texttt{versioned\_quad} table.
          Other operators, such as JoinSQLOperator and GroupSQLOperator, combine these fragments using bitwise operations and other SQL constructs.
    \item \textbf{Finalization:} The FinalizeSQLOperator combines the generated SQL fragments into a single, executable SQL query.
\end{itemize}

\subsection{Sample Dataset}

To illustrate the QuaQue approach, we use a simple versioned RDF dataset about users, their friendships, and their preferences. The dataset consists of three versions, each representing a snapshot of the data at a different point in time.

\begin{table}[h]
    \centering
    \caption{Sample RDF Quads Across Versions}
    \label{tab:sample-dataset}
    \begin{tabular}{lllll}
        \hline
        \textbf{Subject} & \textbf{Predicate} & \textbf{Object} & \textbf{Graph} & \textbf{Versions} \\
        \hline
        :alice           & ex:knows           & :bob            & :g1            & 1,2,3             \\
        :bob             & ex:likes           & "pizza"         & :g1            & 2,3               \\
        :alice           & ex:likes           & "sushi"         & :g1            & 1,3               \\
        :carol           & ex:knows           & :alice          & :g2            & 3                 \\
        :bob             & ex:knows           & :carol          & :g2            & 2,3               \\
        \hline
    \end{tabular}
\end{table}

\begin{figure}
    \centering
    \includegraphics[width=0.95\textwidth]{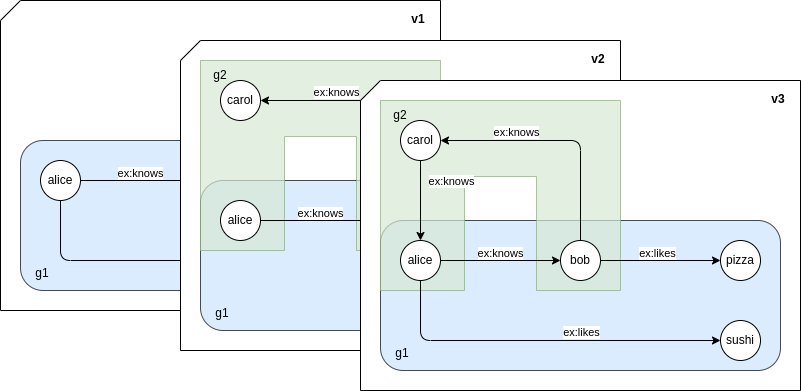}
    \caption{Graphical representation of the sample versioned RDF dataset.}
    \label{fig:sample-dataset}
\end{figure}

Figure~\ref{fig:sample-dataset} provides a graphical representation of the dataset, while Table~\ref{tab:sample-metadata} lists the metadata associated with the versioned graphs.

In the table \ref{tab:sample-metadata}, resources usually prefixed with \texttt{:vng} (e.g., \texttt{:vng1}, \texttt{:vng2}) serve as identifiers for "Versioned Named Graphs", explicitly linking a named graph to a specific version.

\begin{table}[h]
    \centering
    \caption{Metadata of versioned graphs}
    \label{tab:sample-metadata}
    \begin{tabular}{lllll}
        \hline
        \textbf{Subject} & \textbf{Predicate} & \textbf{Object} \\
        \hline
        :vng1            & v:in-version       & 1               \\
        :vng1            & v:version-of       & :g1             \\
        :vng2            & v:in-version       & 2               \\
        :vng2            & v:version-of       & :g1             \\
        :vng3            & v:in-version       & 3               \\
        :vng3            & v:version-of       & :g1             \\
        :vng4            & v:in-version       & 2               \\
        :vng4            & v:version-of       & :g2             \\
        :vng5            & v:in-version       & 3               \\
        :vng5            & v:version-of       & :g2             \\
        \hline
    \end{tabular}
\end{table}

For three versions, the bitstring has three bits (e.g., \texttt{111} for all versions, \texttt{010} for version 2 only).
We illustrate the condensed representation of versioned RDF data in Table~\ref{tab:condensed-dataset}.

\begin{table}[h]
    \centering
    \caption{Condensed Representation of the versioned quads (\texttt{versioned\_quad})}
    \label{tab:condensed-dataset}
    \begin{tabular}{lllll}
        \hline
        \textbf{id\_subj.} & \textbf{id\_pred.} & \textbf{id\_obj.} & \textbf{id\_n\_graph} & \textbf{validity} \\
        \hline
        1                  & 6                  & 2                 & 10                    & 111               \\
        2                  & 7                  & 4                 & 10                    & 011               \\
        1                  & 7                  & 5                 & 10                    & 101               \\
        3                  & 6                  & 1                 & 20                    & 001               \\
        2                  & 6                  & 3                 & 20                    & 011               \\
        \hline
    \end{tabular}
\end{table}

\subsubsection{Quad Pattern Translation}

The translation of a SPARQL quad pattern into SQL in the condensed model is straightforward.
Each quad pattern is mapped to a SQL query over the \texttt{versioned\_quad} table, with conditions on the \texttt{subject}, \texttt{predicate}, \texttt{object}, and \texttt{graph} columns as specified by the quad pattern.
The validity bitstring is always projected, as it encodes the presence of the quad across versions.
Each variable in the quad pattern is represented as a column in the SQL result, prefixed with \texttt{v\$} for variables, \texttt{ng\$} for named graph variables and \texttt{bs\$} for bitstring variables.

\begin{algorithm}
    \caption{Quad pattern translation to SQL}\label{alg:quad-pattern-translation}

    \SetKwProg{Fn}{Function}{ is}{end}

    \KwIn{Quad Pattern $qp$ (from current operator)}
    \KwOut{String $sql\_query$}

    \Fn{TranslateQuadPattern($qp$)}{
        $select \gets ``id\_subject, id\_predicate, id\_object"$\;
        $from \gets ``"$\;
        $where \gets ``"$\;

        \tcc{Check if the quad pattern targets the metadata or a versioned graph}
        \eIf{$qp.graph = ``default graph"$}{
            $from \gets ``metadata"$\;
        } {
            $select \gets select + ", id\_named\_graph, validity"$\;
            $from \gets ``versioned\_quad"$\;
            $where \gets ``bit\_count(validity) <> 0"$\;
        }
        \ForEach{$t$ in $[qp.subject, qp.predicate, qp.object]$}{
            $where \gets where + term\_to\_condition(t)$\;
        }
        \Return{$``SELECT " + select + `` FROM " + from + `` WHERE " + where$}\;
    }
\end{algorithm}

For example, the SPARQL quad pattern:
\begin{lstlisting}[language=SPARQL, caption={Example SPARQL quad pattern.}, label={lst:example-quad-pattern}]
?s <ex:knows> ?o ?g .
\end{lstlisting}

is translated to the following SQL:

\begin{lstlisting}[language=SQL, caption={SQL translation of the example SPARQL quad pattern.}]
SELECT (t0.validity) as bs$g,
    t0.id_subject as v$s,
    t0.id_object as v$o,
    t0.id_named_graph as ng$g
FROM versioned_quad t0
WHERE bit_count(t0.validity) <> 0 AND t0.id_predicate =
    (Subquery to get id of <ex:knows>)
\end{lstlisting}

Given the dataset from Table~\ref{tab:condensed-dataset} and the quad pattern of Query~\ref{lst:example-quad-pattern}, the result of the query is shown in Table~\ref{tab:quad-pattern-result}.

\begin{table}[h]
    \centering
    \caption{Result of the quad pattern query}
    \label{tab:quad-pattern-result}
    \begin{tabular}{llll}
        \hline
        \textbf{bs\$g} & \textbf{v\$s} & \textbf{v\$o} & \textbf{ng\$g} \\
        \hline
        111            & 1             & 2             & 10             \\
        001            & 3             & 1             & 20             \\
        011            & 2             & 3             & 20             \\
        \hline
    \end{tabular}
\end{table}

Variables \texttt{v\$s}, \texttt{v\$o}, and \texttt{ng\$g} correspond to the subject, object, and named graph IDs, and \texttt{bs\$g} is the validity bitstring for each result.

\subsubsection{Join Operation Translation}

The translation of a SPARQL join operation into SQL involves combining the SQL fragments generated for each participating quad pattern.
The key aspect of this translation is the handling of the validity bitstrings.
When two quad patterns are joined, their validity bitstrings are combined using a bitwise AND operation.
This ensures that the resulting rows only include versions where both patterns are valid.

\begin{algorithm}
    \caption{Algorithm of the Join translation}\label{alg:join-translation}
    \SetKwProg{Fn}{Function}{ is}{end}

    \KwIn{Join Operator $join$}
    \KwOut{String $sql\_query$}

    \Fn{TranslateJoin($join$)}{
        $left \gets translate(join.left\_op)$\;
        $right \gets translate(join.right\_op)$\;

        \tcc{Align representations of joined variables}
        \ForEach{$joined\_var$ in $left.vars \cap right.vars$}{
            $l\_var \gets left.vars.get(joined\_var)$\;
            $r\_var \gets right.vars.get(joined\_var)$\;
            \If{$l\_var.repr < r\_var.repr$}{
                $right \gets lower(right, r\_var)$\;
            }
            \If{$r\_var.repr < l\_var.repr$}{
                $left \gets lower(left, l\_var)$\;
            }
        }
        $select \gets get\_joined\_select()$\;
        $from \gets left + ", " + right$\;
        $where \gets get\_joined\_where()$\;
        \Return{$"SELECT " + select + " FROM " + from + " WHERE " + where$}\;
    }
\end{algorithm}

For example, consider the SPARQL join of two quad patterns on a shared graph name variable:

\begin{lstlisting}[language=SPARQL, caption={Example SPARQL join of two quad patterns (join on graph name variable).}, label={lst:example-join-pattern}]
?s <ex:knows> ?o ?g .
?o <ex:likes> ?liked ?g .
\end{lstlisting}

This join between two quad patterns where the joined variables are in a condensed representation is translated into SQL as follows:

\begin{lstlisting}[language=SQL, caption={SQL translation of the example SPARQL join.}]
SELECT (t0.validity & t1.validity) as bs$g, t0.id_subject as v$s, t0.id_named_graph as ng$g, t1.id_object as v$liked, t0.id_object as v$o
FROM versioned_quad t0, versioned_quad t1
WHERE bit_count(t0.validity & t1.validity) <> 0 AND
    t0.id_object = t1.id_subject AND
    t0.id_named_graph = t1.id_named_graph AND
    t0.id_predicate = (Subquery to get id of <ex:knows>) AND
    t1.id_predicate = (Subquery to get id of <ex:likes>)
\end{lstlisting}

Given the dataset in Table~\ref{tab:condensed-dataset}, the result of the join Query~\ref{lst:example-join-pattern} is shown in Table~\ref{tab:join-pattern-result}.

\begin{table}[h]
    \centering
    \caption{Result of the join pattern query}
    \label{tab:join-pattern-result}
    \begin{tabular}{lllll}
        \hline
        \textbf{bs\$g} & \textbf{v\$s} & \textbf{ng\$g} & \textbf{v\$liked} & \textbf{v\$o} \\
        \hline
        011            & 1             & 10             & 4                 & 2             \\
        \hline
    \end{tabular}
\end{table}

Here, \texttt{bs\$g} is the bitwise AND of the validity bitstrings for the joined quads, indicating the versions in which both patterns are valid.

Consider another example where the join is between a condensed variable and a non-condensed variable.

\begin{lstlisting}[language=SPARQL, caption={Example SPARQL join of two quad patterns (join between a condensed and a non-condensed variable).}, label={lst:example-join-pattern-non-condensed}]
?s <ex:knows> ?o ?g .
?g <v:in-version> ?v <ng:Metadata> .
\end{lstlisting}

In this case, one of the quad patterns includes a variable that is not condensed (i.e., it does not have a validity bitstring associated with it).
This scenario requires a different approach to ensure that the join is correctly represented in SQL.
In this situation, we need to first flatten the results of the condensed quad pattern to get the capability to join on the non-condensed variable.
The associated SQL translation is as follows:

\begin{lstlisting}[language=SQL, caption={SQL translation of the example SPARQL join between a condensed and a non-condensed variable.}]
SELECT left_table.v$s, right_table.v$v, left_table.v$g, left_table.v$o
FROM (SELECT flatten_table.v$s, vng.id_versioned_named_graph AS v$g, flatten_table.v$o FROM (Quad pattern 1) flatten_table JOIN versioned_named_graph vng ON flatten_table.ng$g = vng.id_named_graph AND get_bit(flatten_table.bs$g, vng.index_version - 1) = 1) left_table
    JOIN (Quad pattern 2) right_table
        ON left_table.v$g = right_table.v$g;
\end{lstlisting}

Given the dataset from Table~\ref{tab:condensed-dataset} and Table~\ref{tab:metadata}, the result of the join query \ref{lst:example-join-pattern-non-condensed} is:

\begin{table}[h]
    \centering
    \caption{Result of the join pattern query between a condensed and a non-condensed variable}
    \label{tab:join-pattern-result-non-condensed}
    \begin{tabular}{lllll}
        \hline
        \textbf{v\$s} & \textbf{v\$o} & \textbf{v\$g} & \textbf{v\$v} \\
        \hline
        1             & 6             & 20            & 1             \\
        1             & 6             & 21            & 2             \\
        1             & 6             & 22            & 3             \\
        3             & 1             & 24            & 3             \\
        2             & 3             & 23            & 2             \\
        2             & 3             & 24            & 3             \\
        \hline
    \end{tabular}
\end{table}

\subsubsection{Group Operation Translation}

Translating a SPARQL group operation into SQL requires aggregating results according to the specified grouping variables.
The GroupSQLOperator achieves this by producing SQL with a \texttt{GROUP BY} clause for the grouping variables, along with the necessary aggregate functions for the selected variables.
The validity bitstring is incorporated to ensure that the aggregation correctly reflects the versioned nature of the data.

\begin{algorithm}
    \caption{Algorithm of the Group translation}\label{alg:group-translation}
    \SetKwProg{Fn}{Function}{ is}{end}

    \KwIn{Group Operator $gp$}
    \KwOut{String $sql\_query$}

    \Fn{TranslateGroup($gp$)}{
        $subquery \gets translate(gp.sub\_op)$\;
        $select \gets ""$\;
        $from \gets ""$\;
        $group\_by \gets ""$\;
        \ForEach{$var$ in $gp.grouped\_vars$}{
            \If{$var.repr = "condensed"$}{
                $var.repr \gets "id"$\;
                $subquery \gets lower(subquery, var)$\;
            }
            $select \gets select + var.name$\;
        }
        \ForEach{$agg$ in $gp.aggregates$}{
            $select \gets select + translate\_aggregate(agg)$\;
        }
        $from \gets "(" + subquery + ") gb"$\;
        $group\_by \gets get\_group\_by(gp.grouped\_vars)$\;
        \Return{$"SELECT " + select + " FROM " + from + " GROUP BY " + group\_by$}\;
    }
\end{algorithm}

For example, consider the SPARQL query that groups by a variable and counts occurrences:
\begin{lstlisting}[language=SPARQL, caption={Example SPARQL group operation.}, label={lst:example-group-operation}]
SELECT ?o (COUNT(?s) AS ?count)
WHERE {
    ?s <ex:knows> ?o ?g .
}
GROUP BY ?o
\end{lstlisting}

This translation highlights a concept that has been studied in the context of query rewriting with aggregation.
Cohen et al.~\cite{cohen2006rewriting} present methods for rewriting queries with aggregation functions using views, which aligns with our approach~\cite{gil2025condensed}.
In our translation, the aggregation function counts the number of '1's in the validity bitstring, representing the number of versions in which each object occurs.
This group operation is translated into SQL as follows:

\begin{lstlisting}[language=SQL, caption={SQL translation of the example SPARQL group operation.}]
SELECT *, agg0 AS v$count
FROM  (SELECT v$o, SUM(bit_count(bs$g)) AS agg0 FROM (
    Quad Pattern
) gp GROUP BY (v$o)) ext 
\end{lstlisting}

Given the dataset from Table~\ref{tab:condensed-dataset}, the result of the group Query~\ref{lst:example-group-operation} is shown in Table~\ref{tab:group-operation-result}.

\begin{table}[h]
    \centering
    \caption{Result of the group operation query}
    \label{tab:group-operation-result}
    \begin{tabular}{ll}
        \hline
        \textbf{v\$o} & \textbf{agg0} = \textbf{v\$count} \\
        \hline
        2             & 3                                 \\
        1             & 1                                 \\
        3             & 2                                 \\
        \hline
    \end{tabular}
\end{table}

This result demonstrates that the aggregation correctly counts the total occurrences across all versions, leveraging the bitstring representation to efficiently compute version-aware aggregates.

%% file: content/chapters/benchmark.tex
\section{Benchmarks}
\subsection{Benchmark Setup}
To evaluate the performance of QuaQue, we conducted a benchmark comparing our system against Jena, high-performance native RDF triple stores.

The benchmark was conducted on a virtual machine hosted on the PAGODA cloud platform provided by LIRIS~\cite{pagoda2025}, offering a stable, high-performance, and controlled environment to ensure the accuracy and reliability of results.
We leveraged Docker, a containerization platform, to deploy each component of the benchmark in isolated environments.
For evaluation, we used dataset and query workloads from the BEAR benchmarks~\cite{bear2025}, which supply a diverse range of versioned RDF graphs and queries, allowing assessment of system performance across different data sizes.
Utilizing the official BEAR queries ensures our evaluation remains standardized and comparable to previous studies.
A fixed memory limit was applied throughout to maintain consistency and comparability of results.

BEAR archives datasets with different versioning policies, including Time-Based (TB) and Change-Based (CB) versioning.
For this benchmark, we selected the BEAR-B-day dataset, which employs a Time-Based versioning policy.
The BEAR benchmarks focus exclusively on triple pattern and join pattern queries, varying across some index—predicate or predicate and object.

The benchmarking environment used the PAGODA virtual machine provider and Docker as the containerization platform.
OpenNebula was used as the cloud management platform to orchestrate the virtual machines and is supervised by KVM (Kernel-based Virtual Machine) hypervisor with a host passthrough configuration.
The virtual machine ran Ubuntu 24.04 LTS as the operating system with
500GB of allocated disk space.
An AMD EPYC 7443 24-Core Processor (48 threads) @ 2.85 GHz CPUs was utilized for the benchmarking tests.
This environment had access to a total of 12 virtual CPU cores and 64GB of RAM.

\subsection{Development}
For the relational backend of QuaQue, we utilized \textbf{PostgreSQL 15}.
To support efficient query answering for any given triple pattern, we implemented a comprehensive indexing strategy inspired by the Hexastore approach \cite{weiss2008hexastore}.
We created composite B-Tree indexes on six permutations of the quad components (Graph, Subject, Predicate, Object).
This ensures that the query optimizer can utilize an index-only scan for any combination of bound and unbound variables.
The specific index definitions are:
\begin{itemize}
    \item $(id\_named\_graph, id\_subject, id\_predicate, id\_object)$
    \item $(id\_named\_graph, id\_subject, id\_object, id\_predicate)$
    \item $(id\_named\_graph, id\_predicate, id\_object, id\_subject)$
    \item $(id\_named\_graph, id\_predicate, id\_subject, id\_object)$
    \item $(id\_named\_graph, id\_object, id\_predicate, id\_subject)$
    \item $(id\_named\_graph, id\_object, id\_subject, id\_predicate)$
\end{itemize}

Additionally, there is an index on the \texttt{digest} column of the \texttt{resource\_or\_literal} table to speed up lookups of RDF terms.

\subsection{Benchmark Results}

The results of the storage consumption and query performance evaluations are presented below and are published in more detail in a Zenodo repository~\cite{zenodo2025quaque}.

\subsubsection{Storage Efficiency}
Table \ref{tab:benchmark-results-space} compares the disk space usage.
Native RDF stores, such as Jena TDB2, are highly optimized for storage, utilizing dictionary encoding to map URIs (Uniform Resource Identifiers) and literals to integers, resulting in a compact footprint (694 MB).
QuaQue, despite also employing dictionary encoding, exhibits higher storage consumption (4.7 GB) due to its comprehensive indexing strategy on the relational engine.
We also include \textbf{QuaQue-flat} in our comparison, which serves as a baseline relational implementation where each quad-version pair is stored separately.
This is a known trade-off in relational RDF mapping: trading storage space (via exhaustive indexing) for query flexibility and performance.

\begin{table}[h]
    \centering
    \begin{tabular}{lllr}
        \hline
        \textbf{Dataset} & \textbf{Policy} & \textbf{Tool} & \textbf{Space (MB)} \\
        \hline
        BEAR-B-day       & TB              & Jena TDB2     & \textbf{694.39}     \\
        BEAR-B-day       & TB              & QuaQue-flat   & 6489.91             \\
        BEAR-B-day       & TB              & QuaQue        & 4707.63             \\
        \hline
    \end{tabular}
    \caption{Storage consumption comparison.}
    \label{tab:benchmark-results-space}
\end{table}

\subsubsection{Query Execution Time}
Table \ref{tab:benchmark-results-queries} reports the execution times for BEAR-B query templates.
To ensure accurate and stable measurements, each query was executed 200 times.
The first 50 executions were treated as a warm-up phase to mitigate cold-start effects, and the reported results are the average of the final 150 runs.
QuaQue demonstrates better performance, performing better than Jena TDB2 in all observed categories (Join, Predicate-Object, and Predicate queries).
A Mann-Whitney U-test confirmed the statistical significance of the following results.
This suggests that the overhead of the SQL layer is compensated by the efficiency of the PostgreSQL query planner and the availability of covering indexes.

\begin{table*}[h]
    \centering
    \begin{tabular}{lclrrr}
        \hline
        \textbf{Dataset} & \textbf{Policy} & \textbf{Query type} & \textbf{QuaQue}  & \textbf{QuaQue-flat} & \textbf{Jena TDB2} \\
        \hline
        BEAR-B-day       & TB              & Join                & \textbf{6936.00} & 7006.50              & 8096.50            \\
        BEAR-B-day       & TB              & P-O                 & 7309.50          & \textbf{7007.00}     & 7798.50            \\
        BEAR-B-day       & TB              & P                   & \textbf{6533.50} & 6730.50              & 7644.00            \\
        \hline
    \end{tabular}
    \caption{Average query execution times (ms).}
    \label{tab:benchmark-results-queries}
\end{table*}

\begin{figure*}[h!]
    \centering
    \begin{subfigure}[b]{0.47\linewidth}
        \includegraphics[width=\linewidth]{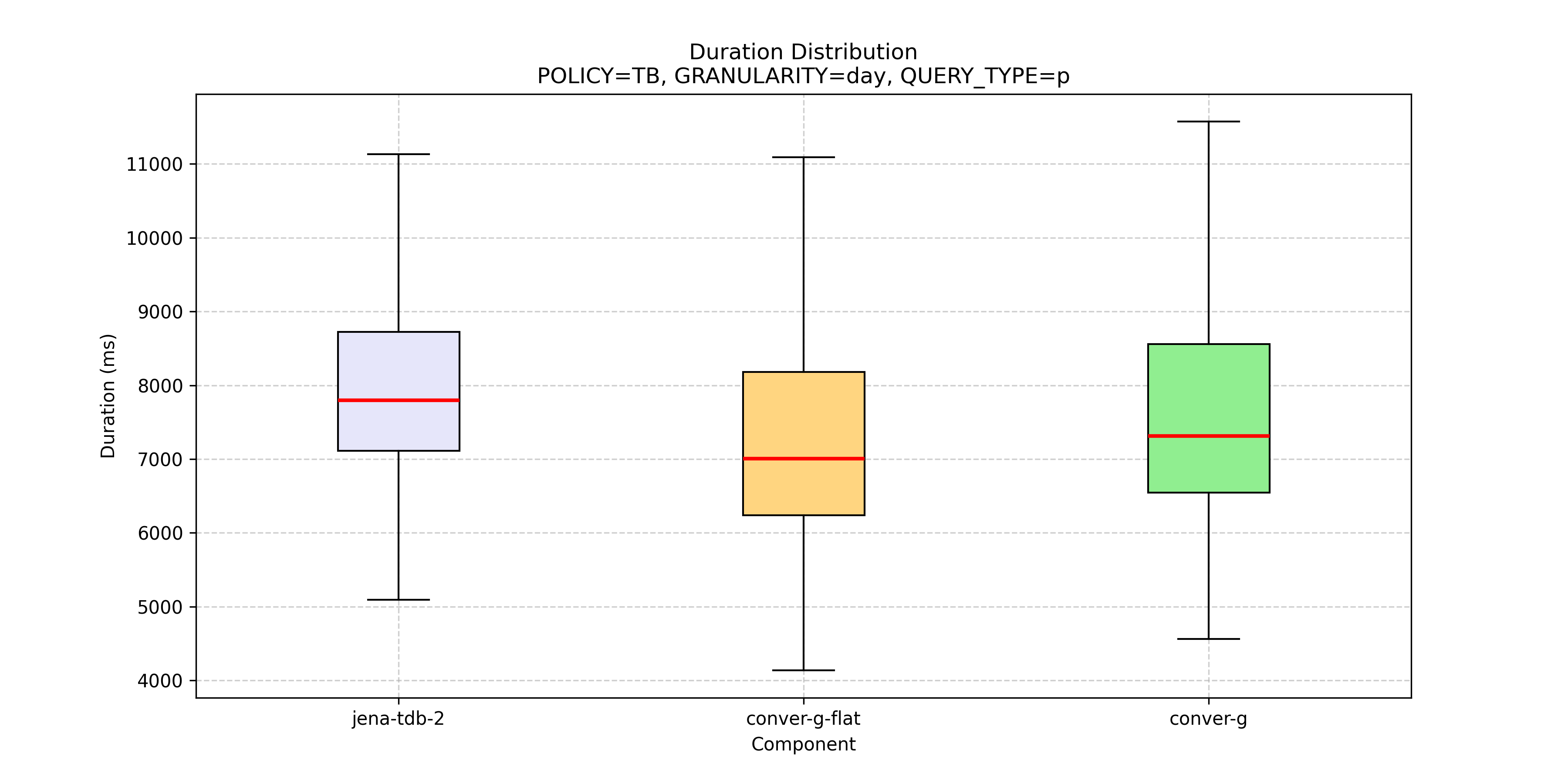}
        \caption{Query Times for predicate index queries}
        \label{fig:benchmark-results-queries-p}
    \end{subfigure}
    \begin{subfigure}[b]{0.47\linewidth}
        \includegraphics[width=\linewidth]{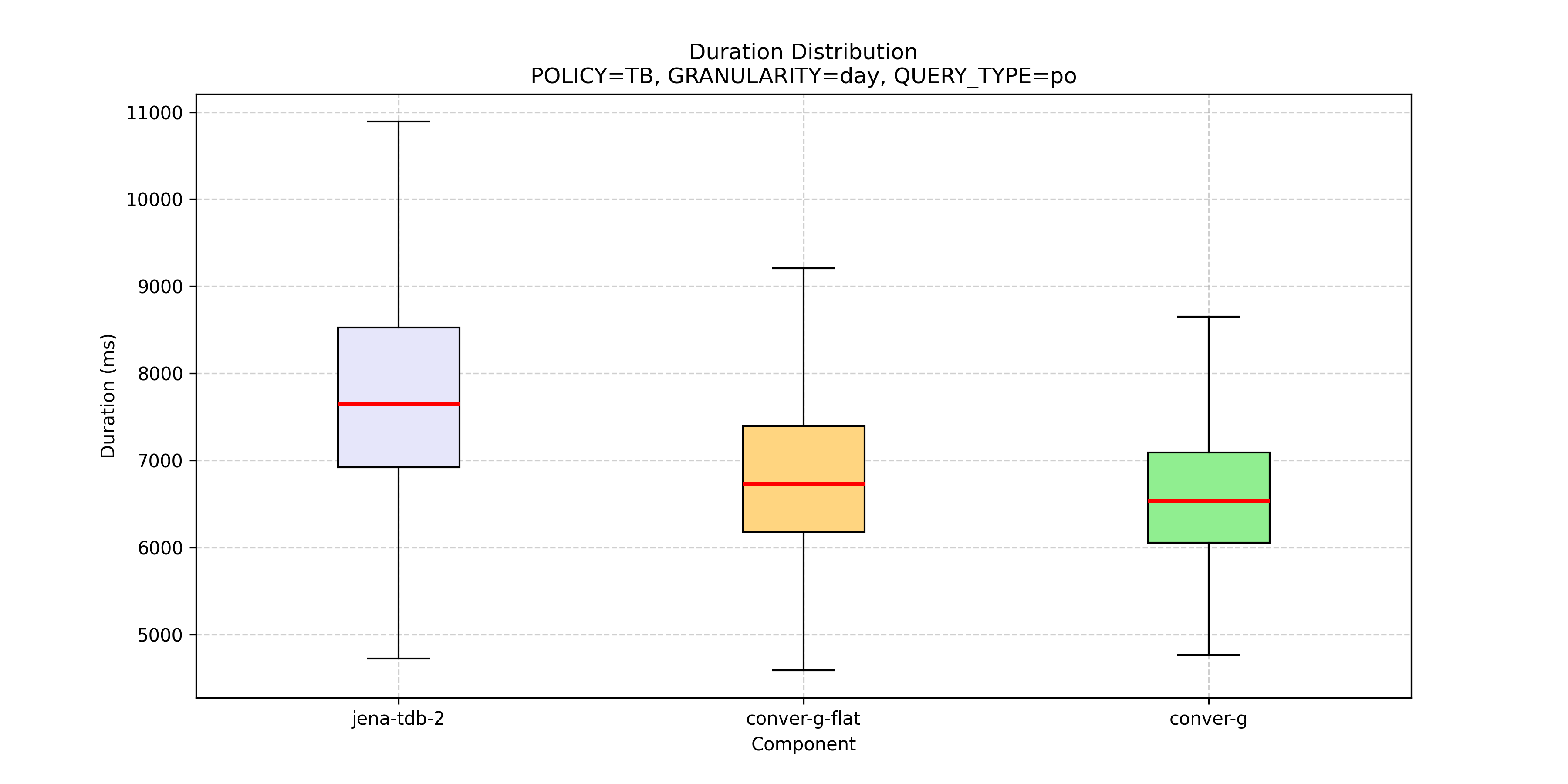}
        \caption{Query Times for predicate-object index queries}
        \label{fig:benchmark-results-queries-po}
    \end{subfigure}
    \begin{subfigure}[b]{0.47\linewidth}
        \includegraphics[width=\linewidth]{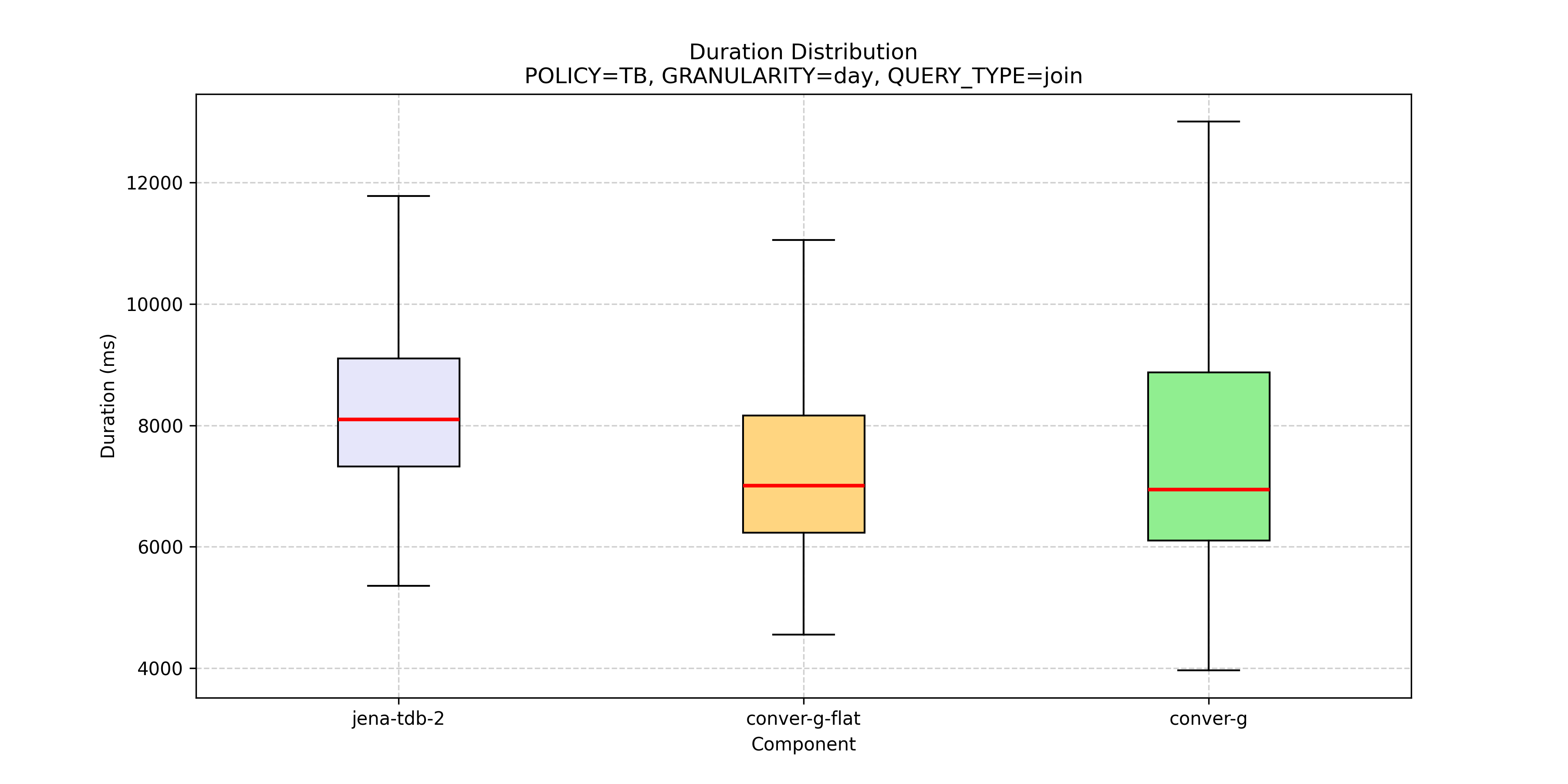}
        \caption{Query Times for join queries}
        \label{fig:benchmark-results-queries-join}
    \end{subfigure}
    \caption{Benchmark Results - Query Times}
    \label{fig:benchmark-results-queries}
\end{figure*}

\subsubsection{Discussion}
The experimental results highlight a trade-off between storage efficiency and query performance that warrants critical analysis.
As shown in Table \ref{tab:benchmark-results-space}, QuaQue's storage footprint is larger than that of Jena TDB2 (4.7 GB vs. 694 MB).
This is a direct consequence of our exhaustive indexing strategy, which maintains six B-Tree indexes to cover all possible quad patterns and the index on the digest of the resource values.

However, this investment in storage yields substantial dividends in query execution time.

Table \ref{tab:benchmark-results-queries} shows that QuaQue achieves improvements of approximately 14\% for predicate queries, 6\% for predicate-object queries, and 14\% for join queries compared to Jena TDB2.
While these gains are statistically significant, a nearly sevenfold increase in storage for a 10--15\% performance improvement represents a trade-off that may not be justified in storage-constrained environments.
Figure \ref{fig:benchmark-results-queries} presents box plots showing the distribution of query execution times.
These plots reveal that QuaQue exhibits lower median times and reduced variability compared to Jena TDB2, which may be valuable in latency-sensitive applications where predictability matters.

These results suggest that the primary value of our approach lies not in raw performance gains over highly optimized native RDF stores, but rather in the flexibility of using standard SQL infrastructure and the potential for integration with existing relational ecosystems.
The condensed relational model can serve as an effective backend for versioned knowledge graph querying when storage capacity is not a constraint and when interoperability with relational systems is a priority.

%% file: content/chapters/conclusion.tex
\section{Conclusion and Future Works}

In this paper, we addressed the challenge of efficiently querying versioned Knowledge Graphs.
We introduced QuaQue, a system that bridges the gap between Semantic Web standards and Relational Database Management Systems.
Our approach relies on a condensed relational model that uses bitstrings to represent the validity of quads across multiple versions, avoiding data redundancy while enabling efficient cross-version querying.

\subsection{Discussions and Future work}
\subsubsection{Benchmark extension}
While our current evaluation provides valuable insights, it is limited to basic query patterns.
Future work will extend the benchmark to include aggregate queries for a more comprehensive assessment of real-world workloads.
Additionally, we plan to compare QuaQue against other versioning policies, such as Change-Based (CB) and Independent Copies (IC), to better understand the trade-offs between storage efficiency and query performance.

\subsubsection{Extensive implementation}
Standard relational algebra lacks support for recursive queries essential for graph analysis, such as path traversal.
To address this, we plan to extend our implementation with advanced operators, such as Agrawal's alpha ($\alpha$)~\cite{agrawal2002alpha}.
This extension will enable efficient execution of complex path-finding and recursive queries, significantly enhancing the system's analytical capabilities.

\subsubsection{Reproducibility}
We emphasize that the approach presented in this paper, implemented in the ConVer-G tool, as well as the benchmark used for evaluation, are fully reproducible.
The source code, datasets, and experimental scripts are publicly available.
To facilitate verification, we provide a containerized environment (Docker) that automates the setup of the database, the loading of datasets, and the execution of the benchmark queries.

In conclusion, QuaQue represents a step towards robust and scalable management of evolving Knowledge Graphs, offering a practical solution for domains requiring complex, concurrent versioning.

%% file: content/acknowledments.tex
\section*{Acknowledgments}

This work, \emph{\papertitle}, was funded by the Universit\'{e} Claude Bernard Lyon 1 as an ATER position, the IADoc@UDL project, and supported by the LIRIS UMR 5205.
We would also like to express our sincere gratitude to the BD team and all members of the Virtual City Project~\cite{vcity2025} for their insightful feedback, constructive discussions, and continuous support throughout the development of this work.
Their expertise and collaboration have been instrumental in shaping the direction and quality of this research.

%% file: content/appendix.tex
\appendix

\section{Appendix}

\subsection{QuaQue: A Queryable Versioned Quad Store}
\subsubsection{Sample Dataset}

The metadata about versions and graphs is represented in Table~\ref{tab:metadata}.

\begin{table}[h]
    \centering
    \caption{Representation of the metadata (\texttt{metadata})}
    \label{tab:metadata}
    \begin{tabular}{lllll}
        \hline
        \textbf{id\_subject} & \textbf{id\_predicate} & \textbf{id\_object} \\
        \hline
        20                   & 8                      & 1                   \\
        20                   & 9                      & 10                  \\
        21                   & 8                      & 2                   \\
        21                   & 9                      & 10                  \\
        22                   & 8                      & 3                   \\
        22                   & 9                      & 10                  \\
        23                   & 8                      & 2                   \\
        23                   & 9                      & 11                  \\
        24                   & 8                      & 3                   \\
        24                   & 9                      & 11                  \\
        \hline
    \end{tabular}
\end{table}

The following Table~\ref{tab:dictionary-dataset} provides the dictionary mapping for resources and literals.

\begin{table}[h]
    \centering
    \caption{Dictionary Representation of the resources or literals (\texttt{resource\_or\_literal})}
    \label{tab:dictionary-dataset}
    \begin{tabular}{lll}
        \hline
        \textbf{id\_resource\_or\_literal} & \textbf{name} & \textbf{type} \\
        \hline
        1                                  & :alice        & resource      \\
        2                                  & :bob          & resource      \\
        3                                  & :carol        & resource      \\
        4                                  & "pizza"       & literal       \\
        5                                  & "sushi"       & literal       \\
        6                                  & ex:knows      & resource      \\
        7                                  & ex:likes      & resource      \\
        8                                  & v:in-version  & resource      \\
        9                                  & v:version-of  & resource      \\
        10                                 & :g1           & resource      \\
        11                                 & :g2           & resource      \\
        20                                 & :vng1         & resource      \\
        21                                 & :vng2         & resource      \\
        22                                 & :vng3         & resource      \\
        23                                 & :vng4         & resource      \\
        24                                 & :vng5         & resource      \\
        \hline
    \end{tabular}
\end{table}